\documentclass[aps,prd,twocolumn,preprintnumbers,
superscriptaddress,showpacs,floatfix]{revtex4}

\usepackage{bm}
\usepackage{epsfig}
\usepackage{graphics}

\newcommand{\eeq}{\end{equation}}
\newcommand{\beq}{\begin{equation}}
\newcommand{\ba}{\begin{array}}
\newcommand{\ea}{\end{array}}
\newcommand{\bea}{\begin{eqnarray}}
\newcommand{\eea}{\end{eqnarray}}

\newcommand{\hepth}[1]{{\tt hep-th/#1}}
\newcommand{\hepph}[1]{{\tt hep-ph/#1}}

\newcommand{\astroph}[1]{{\tt astro-ph/#1}}

\begin{document}

\preprint{UT-STPD-1/07} \preprint{MAN/HEP/2007/3}

\title{Reducing the spectral index in F-term hybrid inflation
through \\ a complementary modular inflation}

\author{G. Lazarides}
\email{lazaride@eng.auth.gr}
\affiliation{Physics Division, School of Technology,
Aristotle University of Thessaloniki,
Thessaloniki 54124, Greece}
\author{C. Pallis}
\email{Constantinos.Pallis@manchester.ac.uk} \affiliation{School
of Physics and Astronomy, The University of Manchester, Manchester
M13 9PL, United Kingdom}


\begin{abstract}
We consider two-stage inflationary models in which a superheavy
scale F-term hybrid inflation is followed by an intermediate scale
modular inflation. We confront these models with the restrictions
on the power spectrum $P_{\cal R}$ of curvature perturbations and
the spectral index $n_{\rm s}$ implied by the recent data within
the power-law cosmological model with cold dark matter and a
cosmological constant. We show that these restrictions can be met
provided that the number of e-foldings $N_{\rm HI*}$ suffered by
the pivot scale $k_*=0.002/{\rm Mpc}$ during hybrid inflation is
appropriately restricted. The additional e-foldings required for
solving the horizon and flatness problems can be naturally
generated by the subsequent modular inflation. For central values
of $P_{\cal R}$ and $n_{\rm s}$, we find that, in the case of
standard hybrid inflation, the values obtained for the grand
unification scale are close to its supersymmetric value
$M_{\rm GUT}=2.86\times10^{16}~{\rm GeV}$, the relevant coupling
constant is relatively large ($\approx 0.005-0.14$), and
$10\lesssim N_{\rm HI*}\lesssim 21.7$. In the case of shifted
[smooth] hybrid inflation, the grand unification scale can be
identified with $M_{\rm GUT}$ provided that
$N_{\rm HI*}\simeq 21$ [$N_{\rm HI*}\simeq 18$].
\end{abstract}

 \maketitle

\section{Introduction} The recently announced three-year results
\cite{wmap3} from the Wilkinson microwave anisotropy probe (WMAP3)
bring  under  considerable stress the well-motivated,
popular, and quite natural models \cite{hsusy} of supersymmetric
(SUSY) F-term hybrid inflation (FHI) \cite{hybrid}, realized
\cite{susyhybrid} at (or close to) the SUSY grand unified theory
(GUT) scale $M_{\rm GUT}=2.86\times10^{16}~{\rm GeV}$. This is due
to the fact that, in these models, the predicted spectral index
$n_{\rm s}$ is too close to unity and without much running. Moreover,
in the presence of non-renormalizable terms generated by supergravity
(SUGRA) corrections with canonical K\"ahler potential, $n_{\rm s}$
approaches~\cite{senoguz} unity more drastically and can even exceed
it. This is in conflict with
the WMAP3 prediction. Indeed, fitting the WMAP3 data with the
standard power-law cosmological model with cold dark matter and a
cosmological constant ($\Lambda$CDM), one obtains \cite{wmap3}
that, at the pivot scale $k_*=0.002/{\rm Mpc}$,
\begin{equation}\label{nswmap}
n_{\rm s}=0.958\pm0.016~\Rightarrow~0.926\lesssim n_{\rm s}
\lesssim 0.99
\end{equation}
at 95$\%$ confidence level.

A way out of this inconsistency is \cite{lofti,king} based on
the utilization of a quasi-canonical K\"ahler potential. With a
convenient arrangement of the signs, a negative mass term can be
induced \cite{king, gpp} in the inflationary potential of the
FHI models. As a consequence, the inflationary path acquires a
local maximum. Under suitable initial conditions, the so-called
hilltop inflation~\cite{lofti} can take place as the inflaton
rolls from this maximum down to smaller values. In this case,
$n_{\rm s}$ can become consistent with Eq.~(\ref{nswmap}), but
only at the cost of an extra indispensable mild tuning \cite{gpp}
of the initial conditions. Alternatively, it is suggested
\cite{battye} that $n_{\rm s}$'s between 0.98 and 1 can be
made compatible with the data by taking into account a
sub-dominant contribution to the curvature perturbation due to
cosmic strings, which may be (but are not necessarily
\cite{trotta}) formed during the phase transition at the end of
FHI. In such a case, the resulting GUT scale is constrained to
values well below the SUSY GUT scale \cite{mairi,jp,gpp}.

In this paper, we propose a two-step inflationary set-up which
allows acceptable $n_{\rm s}$'s in the context of the FHI models
even with canonical K\"ahler potential and without cosmic strings.
The key point in our proposal is that the total
number of e-foldings $N_{\rm tot}$ required for the resolution of
the horizon and flatness problems of the standard big bang
cosmology does not have to be produced exclusively during the GUT
scale FHI. Since $n_{\rm s}$ within the FHI models generally
decreases with the number of e-foldings $N_{\rm HI*}$ that the
pivot scale $k_*$ suffers during FHI, we could constrain $N_{\rm HI*}$
so that Eq.~(\ref{nswmap}) is satisfied. The residual number of
e-foldings $N_{\rm tot}-N_{\rm HI*}$ can be obtained by a second
stage of inflation realized at a lower scale. We call this type of
inflation, which complements the number of e-foldings produced
during the GUT scale inflation, complementary inflation. In our
scenario, modular inflation (MI), which can be easily realized
\cite{modular} by a string axion, plays this role and produces the
required additional number of e-foldings $N_{\rm tot}-N_{\rm HI*}$
with
natural values of the relevant parameters. Such a construction is
also beneficial for MI, since the perturbations of the inflaton
in this model are not sufficiently large to account for the
observations, due to its low inflationary energy scale. As an
extra bonus, the gravitino constraint \cite{gravitino} and the
potential topological defect \cite{kibble} problem of FHI can be
significantly relaxed due to the enormous entropy release
taking place after MI (which naturally assures a low reheat
temperature). However, for the same reason, baryogenesis is made
more difficult but not impossible \cite{benakli} in the context
of a larger scheme with (large) extra dimensions. It is
interesting to note that a constrained $N_{\rm HI*}$ was
previously used in Ref.~\cite{yamaguchi} to achieve a sufficient
running of the spectral index. The additional e-foldings were
provided by new inflation \cite{new}.

Below, we briefly review the basic FHI models (Sec.~\ref{fhim})
and describe the calculation of the relevant inflationary
observables (Sec.~\ref{fhi}). Then, we sketch the main features of
MI (Sec.~\ref{min}) and exhibit the constraints imposed on our
cosmological set-up (Sec.~\ref{cont}). We end up with our
numerical results (Sec.~\ref{num}) and conclusions
(Sec.~\ref{con}).

\section{The FHI models}\label{fhim} The FHI can be realized
\cite{hsusy} adopting one of the superpotentials below:
%
\begin{equation} \label{Whi} W=\left\{\matrix{
\kappa S\left(\bar \Phi\Phi-M^2\right)\hfill   & \mbox{for
standard FHI}, \hfill \cr
\kappa S\left(\bar \Phi\Phi-M^2\right)-S{(\bar \Phi\Phi)^2\over
M_{\rm S}^2}\hfill  &\mbox{for shifted FHI}, \hfill \cr
S\left({(\bar \Phi\Phi)^2\over M_{\rm S}^2}-\mu_{\rm
S}^2\right)\hfill  &\mbox{for smooth FHI}, \hfill \cr}
\right. \end{equation}
where $\bar{\Phi}$, $\Phi$ is a  pair of left handed superfields
belonging to non-trivial conjugate representations of a GUT gauge
group $G$ and reducing its rank by their vacuum expectation values
(VEVs), $S$ is a gauge singlet left handed superfield, $M_{\rm S}
\sim 5\times10^{17}~{\rm GeV}$ is an effective cutoff scale
of the order of the string scale, and the parameters $\kappa$ and
$M,~\mu_{\rm S}~(\sim M_{\rm GUT})$ are made positive by field
redefinitions.

The superpotential for standard FHI in Eq.~(\ref{Whi}) is the most
general renormalizable superpotential consistent with a global
${\rm U}(1)$ R symmetry \cite{susyhybrid} under which
\begin{equation}
\label{Rsym}
S\  \to\ e^{i\alpha}\,S,~\bar\Phi\Phi\ \to\ \bar\Phi\Phi,~W \to\
e^{i\alpha}\, W.
\end{equation}
Including in the superpotential for standard FHI the leading
non-renormalizable term, one obtains the superpotential for
shifted \cite{jean} FHI in Eq.~(\ref{Whi}). The superpotential
for smooth \cite{pana1} FHI is produced by further imposing
an extra $Z_2$ symmetry under which $\Phi\rightarrow -\Phi$ and,
thus, allowing only even powers of the combination
$\bar{\Phi}\Phi$.

From the emerging scalar potential, we can deduce that the
vanishing of the D-terms implies that $\vert\langle\bar{\Phi}
\rangle\vert=\vert\langle\Phi\rangle\vert$, while the vanishing
of the F-terms gives the VEVs of the fields in the SUSY vacuum
(in the case where $\bar{\Phi}$, $\Phi$ are not standard model
(SM) singlets, $\langle\bar{\Phi}\rangle$, $\langle{\Phi}\rangle$
stand for the VEVs of their SM singlet directions). These VEVs
are $\langle S\rangle=0$ and $\vert\langle\bar{\Phi}
\rangle\vert=\vert\langle\Phi\rangle\vert=v_{_G}$ with
\begin{equation} \label{vevs} v_{_G}=\left\{\matrix{
M\hfill   & \mbox{for standard FHI}, \hfill \cr
\frac{M}{\sqrt{2\xi}}\sqrt{1-\sqrt{1-4\xi}}\hfill  &\mbox{for
shifted FHI}, \hfill \cr
\sqrt{\mu_{\rm S}M_{\rm S}}\hfill  &\mbox{for smooth FHI}, \hfill
\cr}
\right. \end{equation}
where $\xi=M^2/\kappa M_{\rm S}^2$ with $1/7.2<\xi<1/4$ \cite{jean}.
As a consequence, $W$ leads to the spontaneous breaking of $G$.
The same superpotential $W$ gives also rise to hybrid inflation.
This is due to the fact that, for large enough values of $|S|$,
there exist flat directions i.e. valleys of local minima of the
classical potential with constant (or almost constant in the case
of smooth FHI) potential energy density. If we call $V_{\rm HI0}$
the dominant contribution to the (inflationary) potential energy
density along these directions, we have\vspace*{-0.cm}
\begin{equation} \label{V0} V_{\rm HI0}=\left\{\matrix{
\kappa^2 M^4\hfill   & \mbox{for standard FHI}, \hfill \cr
\kappa^2 M_\xi^4\hfill  &\mbox{for shifted FHI}, \hfill \cr
\mu_{\rm S}^4\hfill  &\mbox{for smooth FHI}, \hfill \cr}
\right.\end{equation}
with $M_\xi=M\sqrt{1/4\xi-1}$. Inflation can be realized if a
slope along the flat direction (inflationary valley) can be
generated for driving the inflaton towards the vacua. In the cases
of standard \cite{susyhybrid} and shifted \cite{jean} FHI, this
slope can be generated by the SUSY breaking on this valley.
Indeed, $V_{\rm HI0}>0$ breaks SUSY and gives rise to logarithmic
radiative corrections to the potential originating from a mass
splitting in the $\bar{\Phi}$, $\Phi$ supermultiplets. On the other
hand, in the case of smooth \cite{pana1} FHI, the inflationary
valley is not classically flat and, thus, there is no need of
radiative corrections. Introducing the canonically normalized
inflaton field $\sigma=\sqrt{2} \vert S\vert$, the relevant
correction $V_{\rm HIc}$ to the inflationary potential can be
written as follows:

\begin{widetext}
\begin{equation} \label{Vcor} V_{\rm HIc}=\left\{\matrix{
{\kappa^4 M^4 {\sf N}\over 32\pi^2}\left(2 \ln {\kappa^2x M^2
\over
Q^2}+(x+1)^{2}\ln(1+x^{-1})\!+\!(x-1)^{2}\ln(1-x^{-1})\right)\hfill
  & \mbox{for standard FHI}, \hfill \cr
{\kappa^4 M_\xi^4\over 16\pi^2}\left(2 \ln {2\kappa^2x_\xi
M_\xi^2 \over
Q^2}+(x_\xi+1)^{2}\ln(1+x_\xi^{-1})\!+\!(x_\xi-1)^{2}
\ln(1-x_\xi^{-1})\right)\hfill
&\mbox{for shifted FHI,} \hfill \cr
-2\mu_{\rm S}^6M_{\rm S}^2/27\sigma^4\hfill  &\mbox{for smooth
FHI}, \hfill \cr}
\right. \end{equation}
\end{widetext}
where ${\sf N}$ is the dimensionality of the representations
to which $\bar{\Phi}$ and $\Phi$ belong in the case of
standard FHI, $Q$ is a renormalization scale, $x=|S|^2/M^2$,
and $x_\xi=\sigma^2/M^2_\xi$. Although in our work rather
large $\kappa$'s are used in the cases of standard and
shifted FHI, renormalization group effects \cite{espinoza}
remain negligible.

For minimal K\"ahler potential, the leading SUGRA
correction $V_{\rm HIS}$ to the scalar potential along the
inflationary valley reads
\cite{hybrid, senoguz, jp}
\begin{equation} \label{Vsugra}  V_{\rm HIS}=V_{\rm HI0}
{\sigma^4\over8m^4_{\rm P}}, \end{equation}
where $m_{\rm P}\simeq 2.44\times 10^{18}~{\rm GeV}$ is the
reduced Planck scale.

Let us also note that the most important contribution \cite{sstad}
to the inflationary potential from the soft SUSY breaking terms
starts \cite{sstad,jp} playing an important role, in the case of
standard FHI, for $\kappa\lesssim5\times10^{-4}$ and so it remains
negligibly small in our set-up due to the large $\kappa$'s
encountered (see Sec.~\ref{num}). This contribution, in general,
does not have \cite{sstad} a significant effect in the cases of
shifted and smooth FHI too.

All in all, the general form of the potential which drives the
various versions of FHI reads
\beq\label{Vol} V_{\rm HI}=V_{\rm HI0}+V_{\rm HIc}+V_{\rm HIS}.
\eeq

It is worth mentioning that the crucial difference between the
standard and the other two realizations of FHI is that, during
standard FHI, both $\bar{\Phi}$ and $\Phi$ vanish and so the GUT
gauge group $G$ is restored. As a consequence, topological defects
such as strings \cite{jp, mairi, gpp}, monopoles, or domain walls
may be produced \cite{pana1} via the Kibble mechanism
\cite{kibble} during the spontaneous breaking of $G$ at the end of
FHI. This is avoided in the other two cases, since the form of
$W$ allows the existence of non-trivial inflationary valleys along
which $G$ is spontaneously broken (with the appropriate Higgs
fields $\bar\Phi$ and $\Phi$ acquiring non-zero values). Therefore,
no topological defects are produced in these cases.

\section{The dynamics of FHI}\label{fhi} Assuming
(see below) that all the cosmological scales cross outside the
horizon during FHI and are not reprocessed during the subsequent
MI, we can apply the standard calculations (see e.g.
Ref.~\cite{lectures}) for the inflationary observables of FHI.

Namely, the number of e-foldings $N_{\rm HI*}$ that the pivot scale
$k_*$ suffers during FHI can be found from
\begin{equation}
\label{Nefold}
N_{\rm HI*}=\:\frac{1}{m^2_{\rm P}}\;
\int_{\sigma_{\rm f}}^{\sigma_{*}}\, d\sigma\: \frac{V_{\rm HI}}
{V'_{\rm HI}},
\end{equation}
where the prime denotes derivation with respect to (w.r.t.)
$\sigma$, $\sigma_{*}$ is the value of $\sigma$ when the pivot
scale $k_*$ crosses outside the horizon of FHI, and
$\sigma_{\rm f}$ is the value of $\sigma$ at the end of FHI,
which can be found, in the slow-roll approximation, from the
condition
\bea \label{slow} &&~~~{\sf max}\{\epsilon(\sigma_{\rm
f}),|\eta(\sigma_{\rm f})|\}=1,~~\mbox{where}~~\nonumber \\
&&\epsilon\simeq{m^2_{\rm P}\over2}\left(\frac{V'_{\rm HI}}{V_{\rm
HI}}\right)^2~~\mbox{and}~~\eta\simeq m^2_{\rm P}~\frac{V''_{\rm
HI}}{V_{\rm HI}}\cdot \eea
In the cases of standard \cite{susyhybrid} and shifted \cite{jean}
FHI, the end of inflation coincides with the onset of the GUT
phase transition, i.e. the slow-roll conditions are violated
infinitesimally close to the critical point $\sigma_{\rm
c}=\sqrt{2}M$ [$\sigma_{\rm c}=M_\xi$] for standard
[shifted] FHI, where the waterfall regime commences (this is valid
even in the case where the term in Eq.~(\ref{Vsugra}) plays an
important role). On the contrary,  the end of smooth \cite{pana1}
FHI is not abrupt since the inflationary path is stable w.r.t.
variations in $\bar\Phi$, $\Phi$ for all $\sigma$'s and
$\sigma_{\rm f}$ is found from Eq.~(\ref{slow}).

The power spectrum $P_{\cal R}$ of the curvature perturbation can
be calculated at the pivot scale $k_{*}$ by
\begin{equation}  \label{Pr}
P^{1/2}_{\cal R}=\: \frac{1}{2\sqrt{3}\, \pi m^3_{\rm P}}\;
\left.\frac{V_{\rm HI}^{3/2}}{|V'_{\rm
HI}|}\right\vert_{\sigma=\sigma_*}\cdot
\end{equation}
Finally, the spectral index $n_{\rm s}$ and its running
$dn_{\rm s}/d\ln k$ are given by
\bea \label{nS} && n_{\rm s}=\: 1-6\epsilon(\sigma_*)\ +\
2\eta(\sigma_*)~~\mbox{and}~~~~~~~\nonumber\\  &&dn_{\rm s}/d\ln k
=\: 2\left(4\eta(\sigma_*)^2-(n_{\rm s}-1\right)^2)/3
-2\xi(\sigma_*)\  ~~~~~\eea
respectively with $\xi\simeq m_{\rm P}^4~V'_{\rm HI} V'''_{\rm HI}/V^2_{\rm
HI}$.

\section{The Basics of MI \label{min}}

After the gravity mediated soft SUSY breaking, the potential which
can support MI has the form \cite{modular}
\begin{equation}
V_{\rm MI}=V_{\rm MI0}-\frac{1}{2}m_s^2s^2+\dots, \label{Vinf}
\end{equation}
where the ellipsis denotes terms which are expected to stabilize
$V_{\rm MI}$ at \mbox{$s\sim m_{\rm P}$} with $s$ being the
canonically normalized real string axion field. Therefore, in the
above formula, we have
\beq V_{\rm MI0}=v_s(m_{3/2}m_{\rm P})^2~~{\rm and}~~m_s\sim
m_{3/2}, \label{Vm} \eeq
where $m_{3/2}\sim 1~{\rm TeV}$ is the gravitino mass and the
coefficient $v_s$ is of order unity, yielding $V_{\rm
MI0}^{1/4}\simeq3\times 10^{10}~{\rm GeV}$. In this model,
inflation can be of the fast-roll type \cite{fastroll}.
The field evolution is given \cite{fastroll} by
\beq s=s_{\rm i}e^{F_s \Delta N_{\rm MI}}~~{\rm with}~~
F_s\equiv\sqrt{{9\over4}+\left(\frac{m_s}{H_s}\right)^2}-{3\over2}.
\label{Fs} \eeq
Here $s_{\rm i}$ is the initial value of $s$ (i.e. the
value of $s$ at the onset of MI),
$H_s\simeq\sqrt{V_{\rm MI0}}/ \sqrt{3}m_{\rm P}$ is the
Hubble parameter corresponding to $V_{\rm MI0}$, and $\Delta
N_{\rm MI}$ is the number of e-foldings obtained from $s=
s_{\rm i}$ until a given $s$.

\par
From Eq.~(\ref{Fs}), we can estimate the total number of
e-foldings during MI as
\beq N_{\rm MI}\simeq\frac{1}{F_s} \ln\left(\frac{s_{\rm f}}
{s_{\rm i}}\right), \label{Nmp} \eeq
where $s_{\rm f}$ is the final value of $s$. This value is
given by $s_{\rm f}={\sf min}\{\langle s\rangle,s_{\rm sr}\}$,
where $\langle s\rangle\sim m_{\rm P}$ is the VEV of $s$
and $s_{\rm sr}$ is determined by the condition
\begin{equation}\label{varepsilon}
\epsilon_{\rm MI}=1~~{\rm with}~~
\epsilon_{\rm MI}\equiv -\frac{\dot{H}_{\rm MI}}{H^2_{\rm MI}}
\simeq\frac{1}{2}F_s^2\left(\frac{s}{m_{\rm P}}\right)^2
\end{equation}
being the slow-roll parameter for MI ($H_{\rm MI}$ is the
Hubble parameter during MI and the dot denotes
derivation w.r.t. the cosmic time). To derive
Eq.~(\ref{varepsilon}), we use the equation of motion for $s$
during MI and Eq.~(\ref{Fs}). For definiteness, we
take $\langle s\rangle=m_{\rm P}$ in our calculation.

\begin{figure*}[t!]
\centering
\includegraphics[width=50mm,angle=-90]{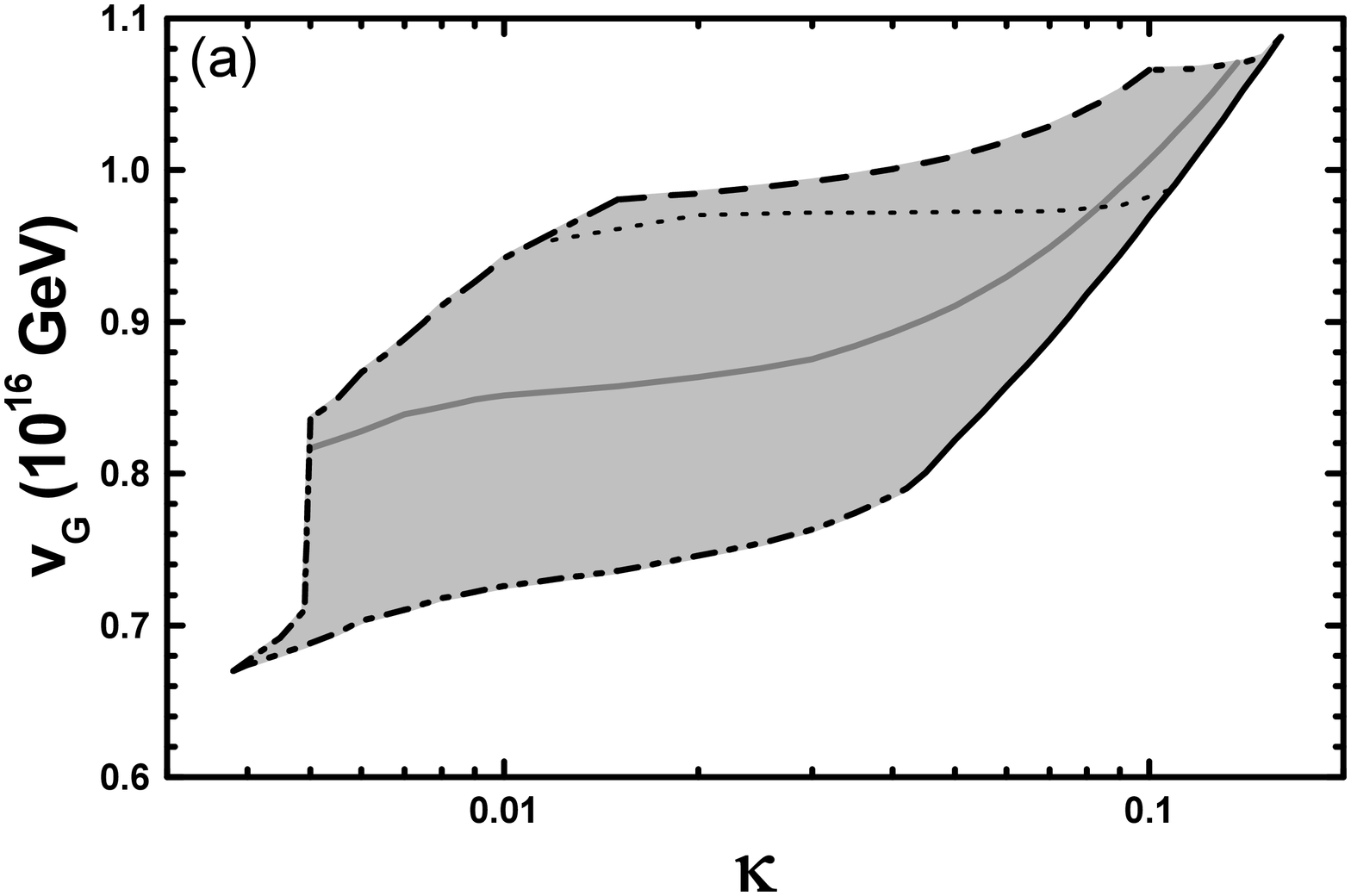}
\includegraphics[width=50mm,angle=-90]{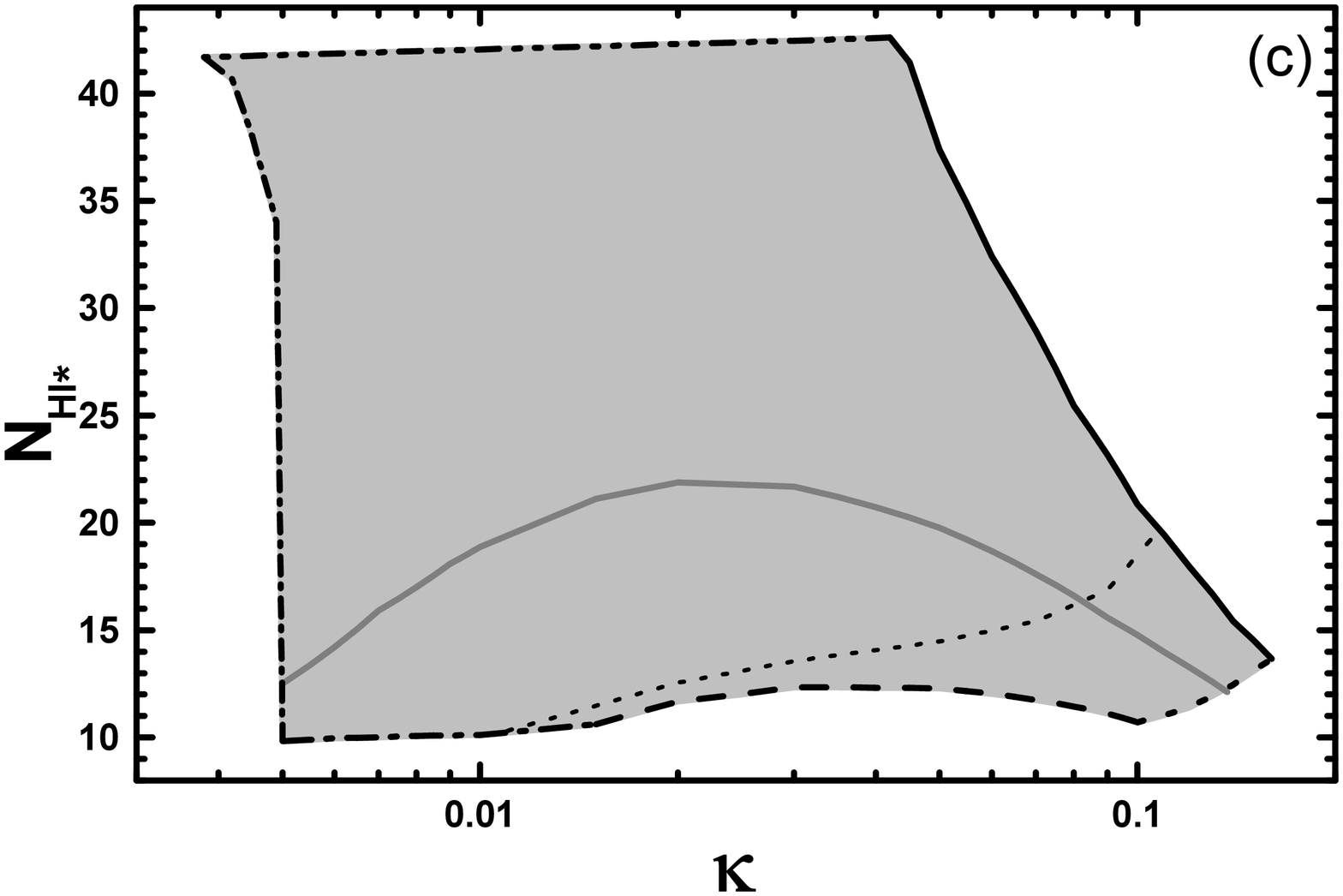}
\centering
\includegraphics[width=50mm,angle=-90]{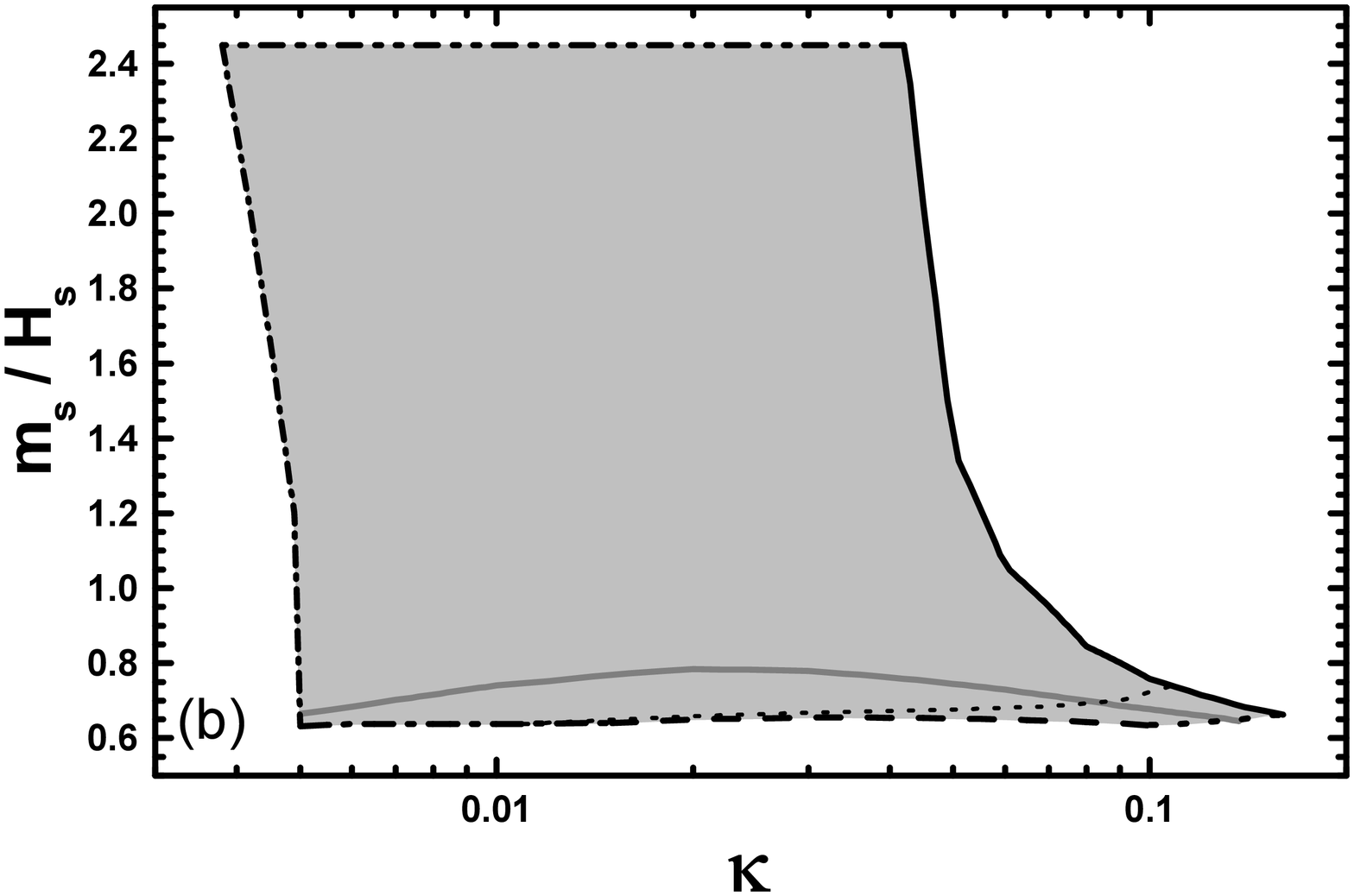}
\includegraphics[width=50mm,angle=-90]{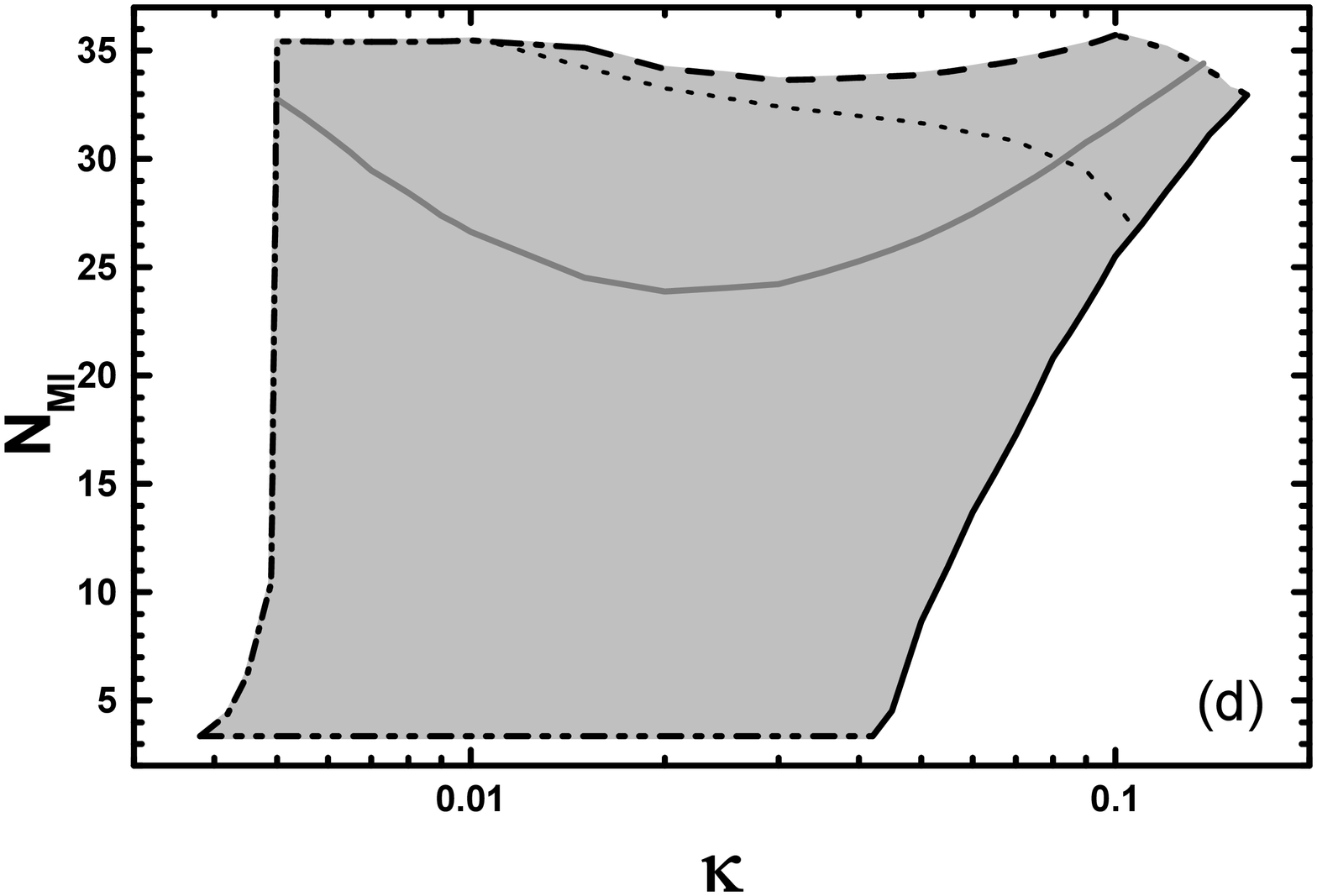}
\caption{\label{stad} Allowed (lightly gray shaded) regions in the
{(a)} $\kappa-v_{_{G}}$, {(b)} $\kappa-m_s/H_s$, {(c)}
$\kappa-N_{\rm HI*}$, and {(d)} $\kappa-N_{\rm MI}$ plane for
standard FHI. The black solid [dashed] lines correspond to the
upper [lower] bound on $n_{\rm s}$ in Eq.~(\ref{nswmap}), whereas
the gray solid lines have been obtained by fixing $n_{\rm s}$ to
its central value in Eq.~(\ref{nswmap}). The dot-dashed
[double dot-dashed] lines correspond to the lower [upper] bound on
$N_{\rm HI*}~[m_s/H_s]$ from Eq.~(\ref{ten}) [Eq. (\ref{nin})].
The bold [faint] dotted lines correspond to
$dn_{\rm s}/d\ln k=-0.01$ [$dn_{\rm s}/d\ln k=-0.005$]. Finally,
the short dash-dotted lines correspond to the lower bound on
$V_{\rm HI0}$ from Eq.~(\ref{random}). In the allowed regions,
Eqs.~(\ref{Prob}) and (\ref{Ntott}) are also satisfied.}
\end{figure*}

\section{Observational Constraints}\label{cont}

The cosmological scenario under consideration needs to satisfy a
number of constraints. These can be outlined as follows:

\begin{itemize}

\item[{\bf (a)}]
The power spectrum in Eq.~(\ref{Pr}) is to be
confronted with the WMAP3 data \cite{wmap3}
\begin{equation}  \label{Prob}
~~~~P^{1/2}_{\cal R}\simeq\: 4.86\times
10^{-5}~~\mbox{at}~~k_*=0.002/{\rm Mpc}.
\end{equation}
\item[{\bf (b)}]
According to the inflationary paradigm, the horizon and flatness
problems of the standard big bang cosmology can be successfully
resolved provided that the pivot scale $k_*$ suffers a certain total
number of e-foldings $N_{\rm tot}$, which depends on some details
of the cosmological scenario. In our set-up, $N_{\rm tot}$
consists of two contributions:
\beq N_{\rm tot} =\: N_{\rm HI*}+N_{\rm MI}\,.\label{Ntot}\eeq
Employing standard methods \cite{hybrid, anupam}, we can easily
derive, in our case, the required $N_{\rm tot}$:
\beq~~~~ N_{\rm tot}\simeq22.6+{2\over 3}\ln{V^{1/4}_{\rm
HI0}\over{1~{\rm GeV}}}+ {1\over3}\ln {T_{\rm Mrh}\over{1~{\rm
GeV}}},\label{Ntott} \eeq
where $T_{\rm Mrh}$ is the reheat temperature after the completion
of MI. Here, we have assumed that the reheat temperature after
FHI is lower than $V_{\rm MI0}^{1/4}$ (as in the majority of these
models \cite{senoguz}) and, therefore, the whole inter-inflationary
period is matter dominated.

\begin{figure*}[t!]
\centering
\includegraphics[width=50mm,angle=-90]{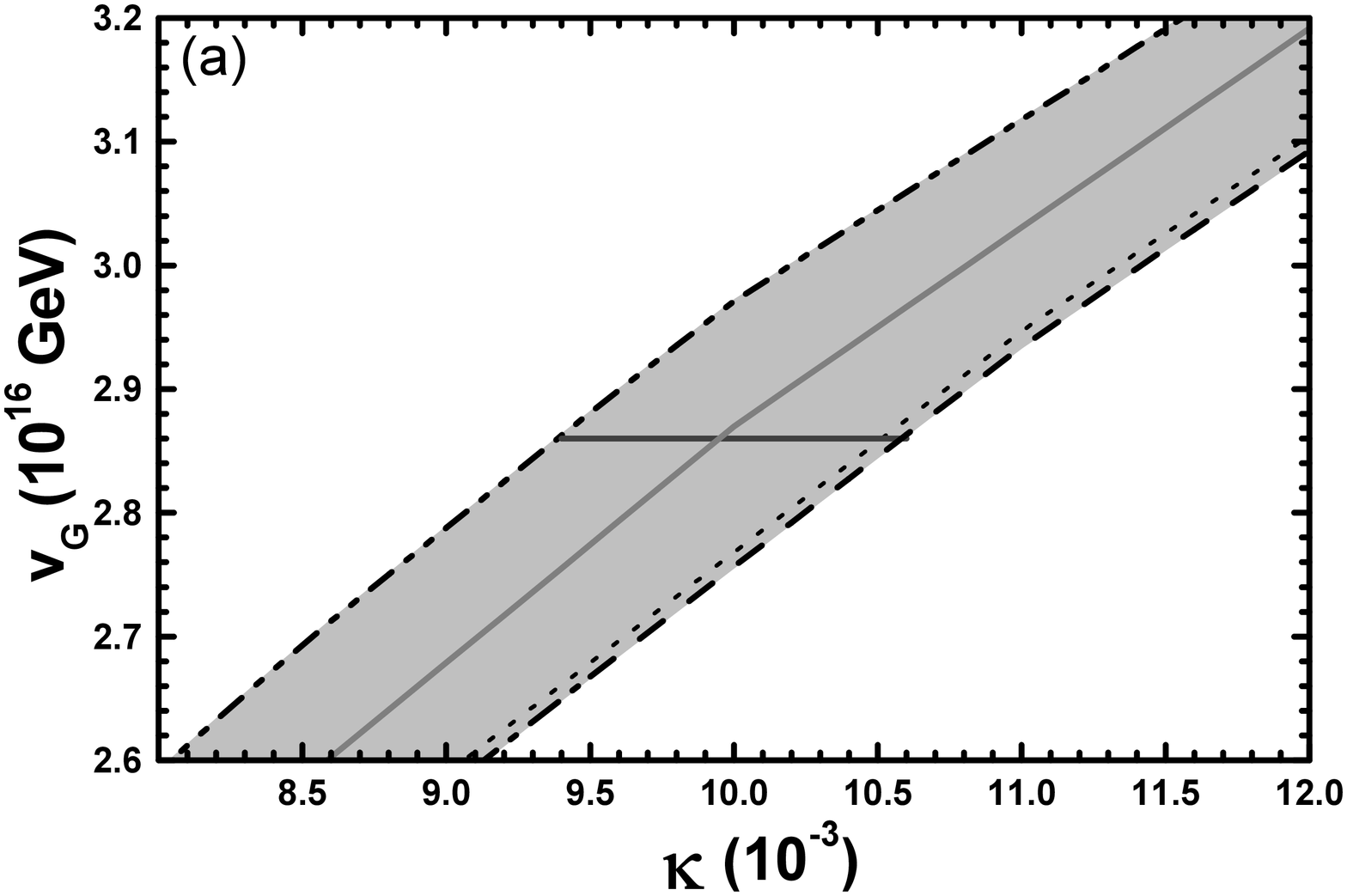}
\includegraphics[width=50mm,angle=-90]{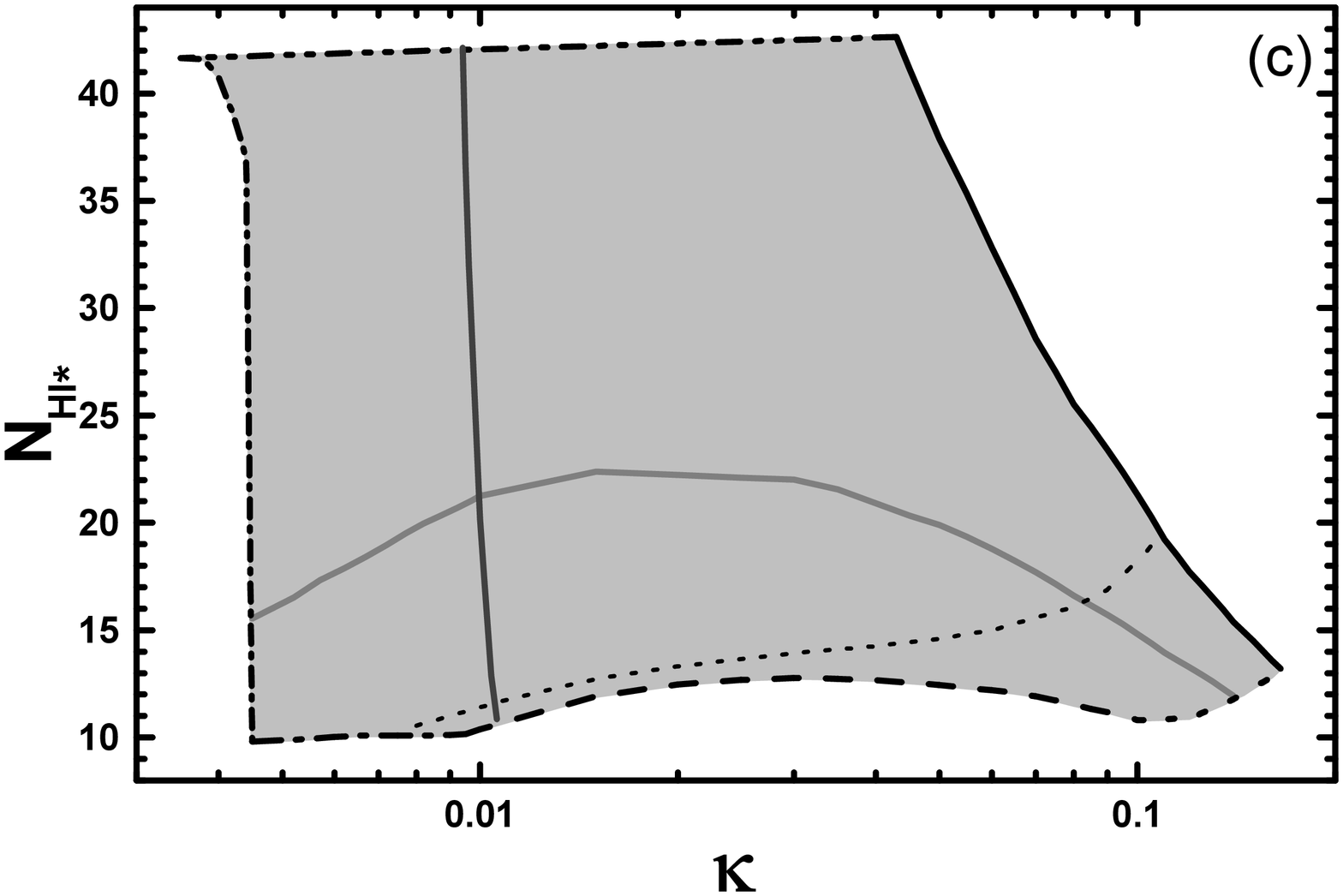}
\centering
\includegraphics[width=50mm,angle=-90]{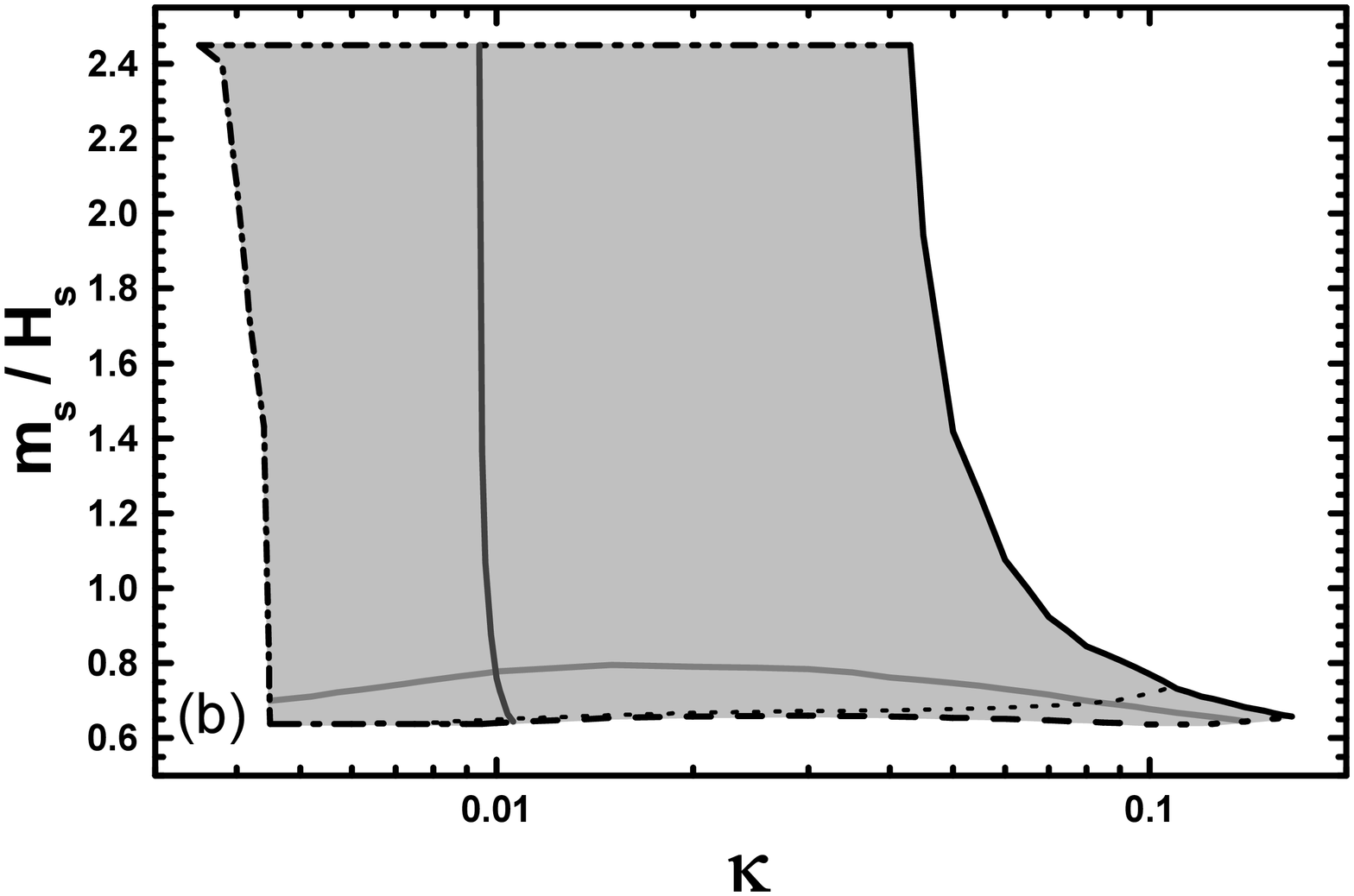}
\includegraphics[width=50mm,angle=-90]{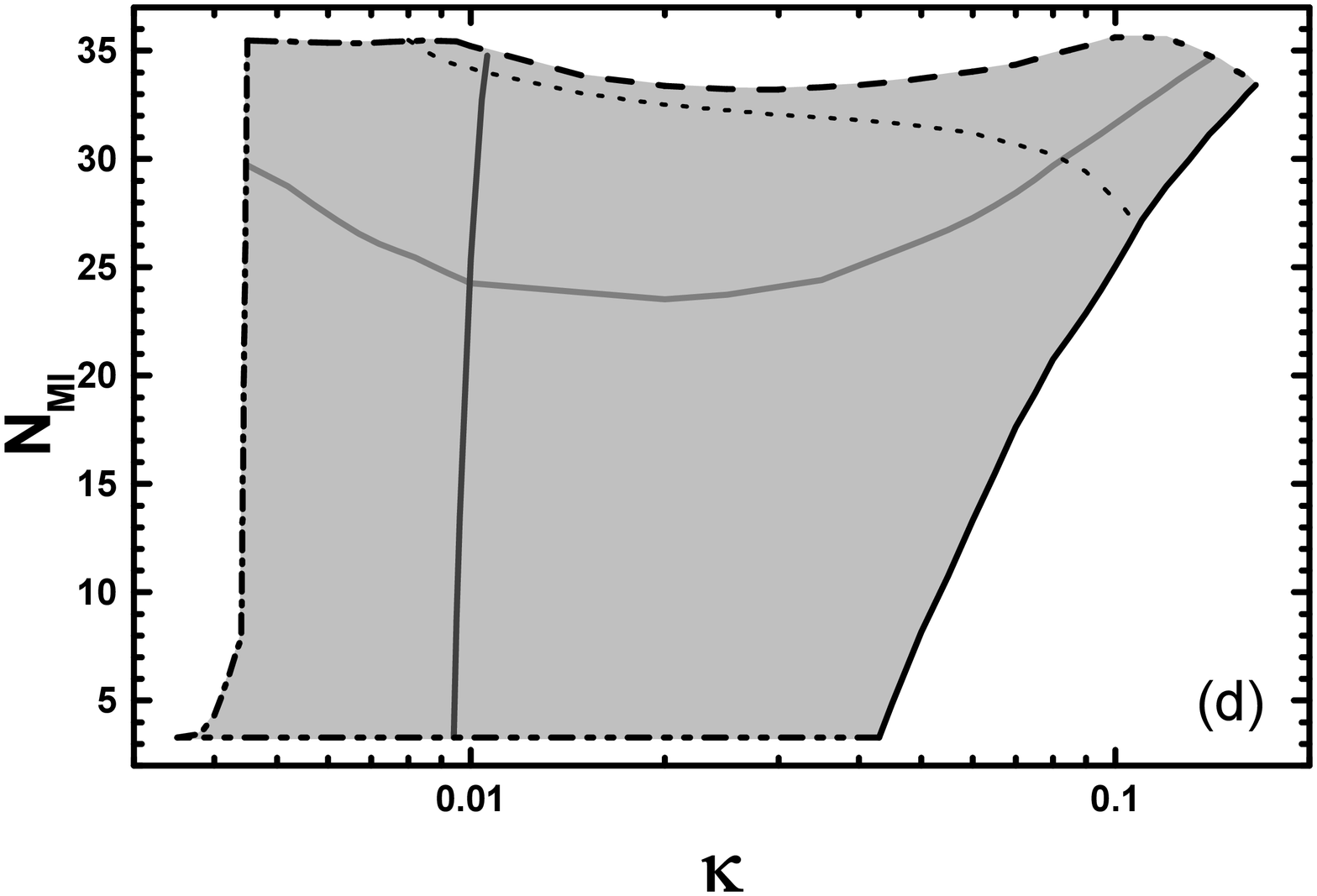}
\caption{\label{shth} Allowed regions in the {(a)}
$\kappa-v_{_G}$, {(b)} $\kappa-m_s/H_s$, {(c)}
$\kappa-N_{\rm HI*}$, and {(d)} $\kappa-N_{\rm MI}$ plane for
shifted FHI with
$M_{\rm S}=5\times10^{17}~{\rm GeV}$. The notation is the
same as in Fig.~\ref{stad}. We also include dark gray solid
lines corresponding to $v_{_{G}}=M_{\rm GUT}$.}
\end{figure*}

\item[{\bf (c)}]
We have also to assure that all the cosmological scales (i)
leave the horizon during FHI and (ii) do not re-enter the horizon
before the onset of MI (this would be possible since the scale
factor increases faster than the horizon during the
inter-inflationary era \cite{anupam}). Both these requirements
can be met if we demand
\cite{anupam,astro} that
\beq N_{\rm HI*}\gtrsim N^{\rm min}_{\rm HI*}\simeq3.9+{1\over
6}\ln {V_{\rm HI0}\over V_{\rm MI0}}\cdot\label{ten}\eeq
%
The first term in the expression for $N^{\rm min}_{\rm HI*}$ is
the number of e-foldings elapsed between the horizon crossing of
the pivot scale $k_*$ and the scale $0.1/{\rm Mpc}$ during FHI.
Note that length scales of the order of $10~{\rm Mpc}$ are
starting to feel nonlinear effects and it is, thus, difficult to
constrain \cite{astro} primordial density fluctuations on smaller
scales. Given that $(V_{\rm HI0}/V_{\rm MI0})^{1/4}\sim
10^{14}/10^{10}\sim 10^4$, we expect that
$N^{\rm min}_{\rm HI*}\sim 10$.

\item[{\bf (d)}]
As it is well known \cite{espinoza}, in the models under
consideration, $|dn_{\rm s}/d\ln k|$ increases as $N_{\rm HI*}$
decreases. Therefore, limiting ourselves to $|dn_{\rm s}/d\ln k|$'s
consistent with the assumptions of the power-law $\Lambda$CDM
cosmological model, we obtain a lower bound on $N_{\rm HI*}$.
Since, within the
cosmological models with running spectral index, $|dn_{\rm s}/d\ln
k|$'s of order 0.01 are encountered \cite{wmap3}, we impose the
following upper bound on $|dn_{\rm s}/d\ln k|$:
\beq \vert dn_{\rm s}/d\ln k\vert \ll 0.01\,. \label{dnsk} \eeq
In our numerical investigation (see Sec.~\ref{num}), we display
boundary curves for $dn_{\rm s}/d\ln k=-0.005$ and $-0.01$.

\item[{\bf (e)}]
For MI to be natural, we constrain the dimensionless
parameter $v_s$ in Eq.~(\ref{Vm}) as follows:
\beq\label{nin}
~~0.5\leq v_s\leq 10~~\Rightarrow~~2.45\gtrsim
m_s/H_s\gtrsim 0.55,
\eeq
where we take $m_s=m_{3/2}$ (see below). The lower bound on $v_s$
is chosen so that the sum of the two explicitly displayed
terms in the right hand side of Eq.~(\ref{Vinf}) is positive
for $s<m_{\rm P}$. From Eq.~(\ref{varepsilon}), we see
that, for the values of $m_s/H_s$ in Eq.~(\ref{nin}),
$s_{\rm sr}>m_{\rm P}$ and, thus, $s_{\rm f}=m_{\rm P}$. Using
Eq.~(\ref{Nmp}), we then find
that the upper bound on $m_s/H_s$ implies the constraint $N_{\rm MI}
\gtrsim 0.73\ln (m_{\rm P}/s_{\rm i})$. Note, though, that
Eqs.~(\ref{Fs})--(\ref{varepsilon}) are not very accurate near the
upper bound on $m_s/H_s$ since, in this region, the slow-roll
parameter $\epsilon_{\rm MI}$ gets too close to unity at
$s=m_{\rm P}$ and,
thus, the Hubble parameter does not remain constant as $s$
approaches $m_{\rm P}$. So our results at large values of
$m_s/H_s$ should be considered only as indicative. Fortunately, as
we will see below, the interesting solutions are found near the
lower bound on $m_s/H_s$, where the accuracy of these formulas is
much better (of the order of a few per cent for $s_{\rm i}\sim
0.01\,m_{\rm P}$). Moreover, the slow-roll parameter for MI
\begin{equation}
\eta_{\rm MI}\equiv m_{\rm P}^2\frac{V_{\rm MI}^{(2)}}{V_{\rm MI}}
\simeq -\frac{1}{3}\left(\frac{m_s}{H_s}\right)^2,
\end{equation}
where we again take $m_s=m_{3/2}$, satisfies the inequality
$|\eta_{\rm MI}|\leq 1$ for $m_s/H_s\lesssim 1.73$ (the
superscript $(n)$ denotes the $n$th derivative w.r.t. the
string axion $s$). So the interesting
solutions correspond to slow- rather than fast-roll MI.
We should also point out that the presence of the
(unspecified) terms in the ellipsis in the right hand side of
Eq.~(\ref{Vinf}), which are needed for stabilizing the potential
at $s\sim m_{\rm P}$, also generates an uncertainty in
Eqs.~(\ref{Fs})--(\ref{varepsilon}). We assume that this
uncertainty is small and neglect it.

\item[{\bf (f)}]
Finally, we assume that FHI lasts long enough so that the
value of the almost massless string axion $s$ is completely
randomized \cite{randomize} as a consequence of its quantum
fluctuations from FHI. We further assume that
\beq\label{random}
V_{\rm MI0}\lesssim H^4_{\rm HI0},
\eeq
where $H_{\rm HI0}=\sqrt{V_{\rm HI0}}/\sqrt{3}m_{\rm P}$ is
the Hubble parameter corresponding to $V_{\rm HI0}$, so that
all the values of $s$ belong to the randomization region
\cite{randomize}. The field $s$ remains practically
frozen during the inter-inflationary period since the
Hubble parameter is larger than its mass. Under these
circumstances, all the initial values $s_{\rm i}$ of $s$
from zero to $m_{\rm P}$ are equally probable. However, we
take $s_{\rm i}\gg H_{\rm HI0}/2\pi$ so that the homogeneity
of our present universe is not jeopardized by the quantum
fluctuations of $s$ from FHI. Note that randomization of
the value of a scalar field via inflationary quantum
fluctuations requires that this field remains almost
massless during inflation. For this, it is important that
the field does not acquire \cite{hybrid, effmass} mass of
the order of the Hubble parameter via the SUGRA scalar
potential. This is, indeed, the case for the string axion
during FHI (and the subsequent inter-inflationary era). In
the opposite case, this field could decrease to very small
values until the onset of MI as the inflaton of new
inflation \cite{new} in Refs.~\cite{yamaguchi,kawasaki}.

\end{itemize}

\section{Numerical results}\label{num}

In the case of standard FHI, we take ${\sf N}=2$. This
corresponds to the left-right symmetric GUT gauge group
${\rm SU(3)_c\times SU(2)_L\times SU(2)_R
\times U(1)_{B-L}}$ with $\bar\Phi$ and $\Phi$ belonging
to ${\rm SU(2)_R}$ doublets with $B-L=-1$ and $1$ respectively.
It is known \cite{trotta} that no cosmic strings are
produced during this realization of standard FHI. As a
consequence, we are not obliged to impose extra
restrictions on the parameters (as e.g. in
Refs.~\cite{mairi,jp}). Let us mention, in passing, that,
in the case of shifted \cite{jean} FHI, the GUT gauge
group is the Pati-Salam group ${\rm SU(4)_c\times SU(2)_L
\times SU(2)_R}$. We take $T_{\rm Mrh}=1~{\rm GeV}$ and
$m_{3/2}=m_s=1~{\rm TeV}$ throughout. These are
indicative values which do not affect crucially our
results. Indeed, $T_{\rm Mrh}$ appears in Eq.~(\ref{Ntott})
through its logarithm and so its variation has a minor
influence on the value of $N_{\rm tot}$. Furthermore,
$N_{\rm MI}$ depends crucially only on $F_s$ -- see
Eq.~(\ref{Nmp}) -- which in turn depends on the ratio
$m_s/H_s$ and not separately on $m_s$ or $H_s$. Finally,
we choose the initial value $s_{\rm i}$ of the string
axion $s$ at the onset of MI to be given by $s_{\rm i}
=0.01\,m_{\rm P}$ in all the cases that we consider. This
value is close enough to $m_{\rm P}$ to have a
non-negligible probability to be achieved by the
randomization of $s$ during FHI (see point {\bf (f)} in
Sec.~\ref{cont}). At the same time, it is adequately
smaller than $m_{\rm P}$ to guarantee good accuracy of
Eqs.~(\ref{Fs})--(\ref{varepsilon}) near the interesting
solutions and justify the fact that we neglect the
uncertainty from the terms in the ellipsis in
Eq.~(\ref{Vinf}) (see point {\bf (e)} in Sec.~\ref{cont}).
Moreover, larger $s_{\rm i}$'s lead to smaller parameter
space for interesting solutions (with $n_s$ near its
central value).

\begin{table}[t]
\caption{Convention for the various lines in
Figs.~\ref{stad}$-$\ref{smth}.}
\begin{tabular}{l@{\hspace{0.cm}}r}
\toprule
Type of Line & Corresponding Condition \\ \colrule\colrule
Black Solid & Upper bound on $n_{\rm s}$ in Eq.~(\ref{nswmap}) \\
\colrule
Dashed &  Lower bound on $n_{\rm s}$ in Eq.~(\ref{nswmap})
\\ \colrule
Short Dash-dotted &  Lower bound on $V_{\rm HI0}$ from
Eq.~(\ref{random})
\\ \colrule
Bold Dotted &  $dn_{\rm s}/d\ln k=-0.01$\\ \colrule
Faint Dotted & $dn_{\rm s}/d\ln k=-0.005$\\ \colrule
Dot-dashed &  Lower bound on $N_{\rm HI*}$ in Eq.~(\ref{ten})\\
\colrule
%
%
Double Dot-dashed & ~~~~Upper bound on $m_s/H_s$ in
Eq.~(\ref{nin})\\ \colrule
Gray Solid & Central value of $n_{\rm s}$ in Eq.~(\ref{nswmap})\\
\colrule
Dark Gray Solid & $v_{_{G}}=M_{\rm GUT}=2.86\times10^{16}~{\rm
GeV}$\\ \botrule
\end{tabular}
\label{tabc}
\end{table}

In our numerical computation, we use, as input parameters,
$\kappa$ (for standard and shifted FHI with fixed $M_{\rm S}=5
\times10^{17}~{\rm GeV}$) or $M_{\rm S}$ (for smooth FHI) and
$\sigma_*$. Using Eqs.~(\ref{nS}) and (\ref{Prob}), we extract
$n_{\rm s}$ and $v_{_G}$ respectively. For every chosen $\kappa$ or
$M_{\rm S}$, we then restrict $\sigma_*$ so as to achieve
$n_{\rm s}$ in the range of Eq.~(\ref{nswmap}) and take the output
values
of $N_{\rm HI*}$ (contrary to the conventional strategy -- see
e.g. Refs.~\cite{sstad, jp} -- in which $N_{\rm HI*}\simeq53$ is
treated as a constraint and $n_{\rm s}$ is an output parameter).
Finally, we find, from Eqs.~(\ref{Ntot}) and (\ref{Ntott}), the
required $N_{\rm MI}$ and the corresponding $v_s$ or $m_s/H_s$
from Eq.~(\ref{Nmp}).

\begin{figure*}[t!]
\centering
\includegraphics[width=50mm,angle=-90]{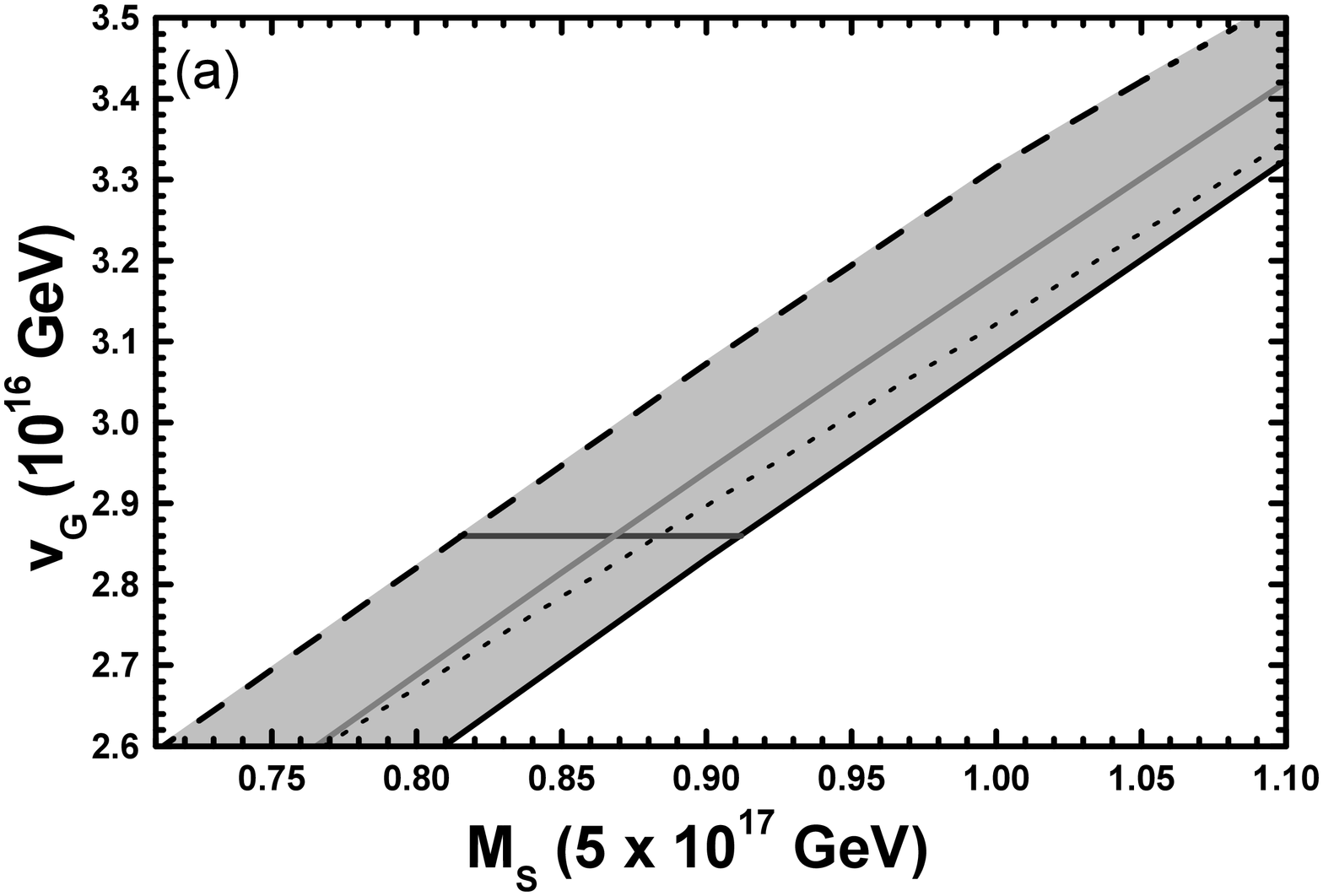}
\includegraphics[width=50mm,angle=-90]{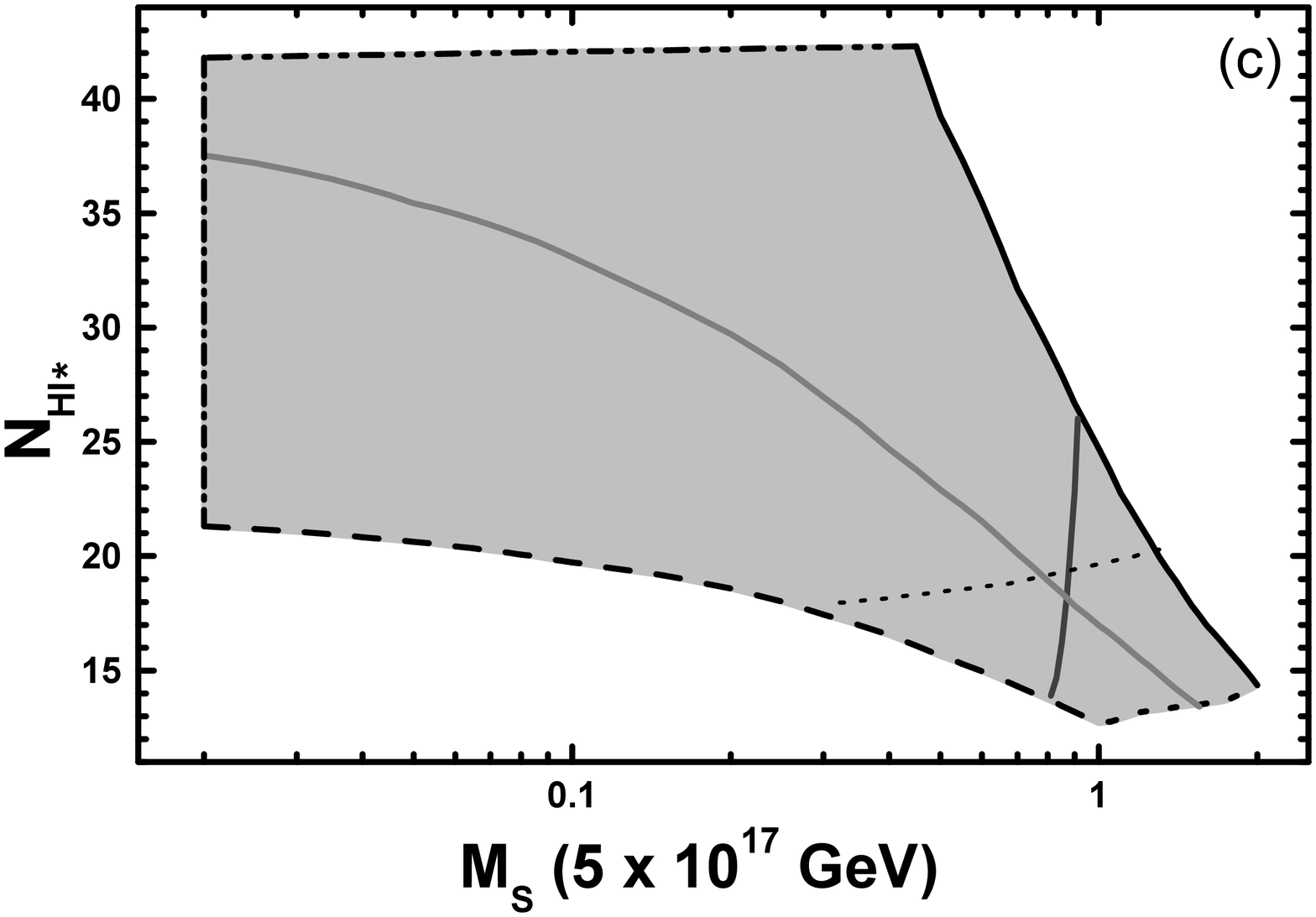}
\centering
\includegraphics[width=50mm,angle=-90]{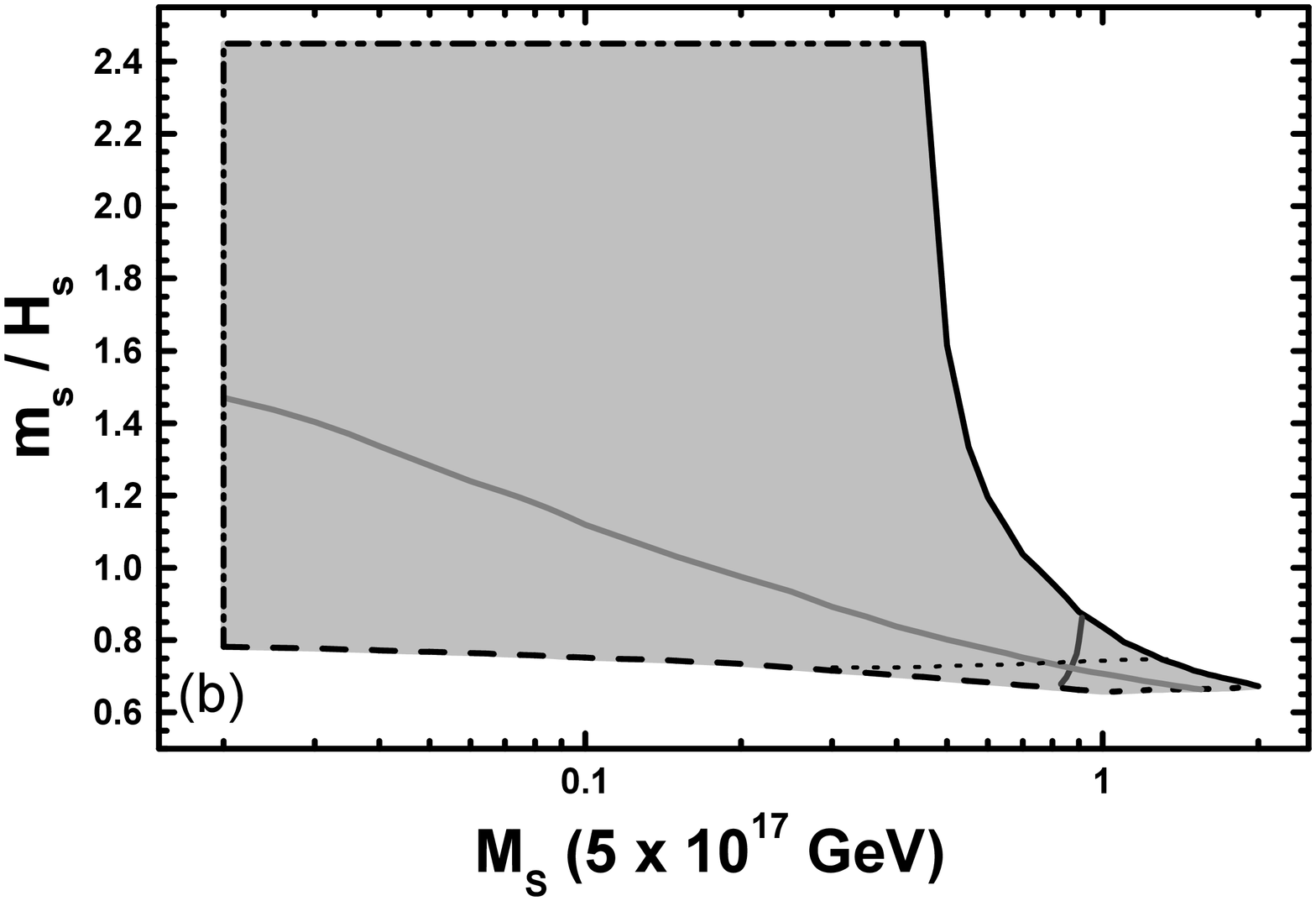}
\includegraphics[width=50mm,angle=-90]{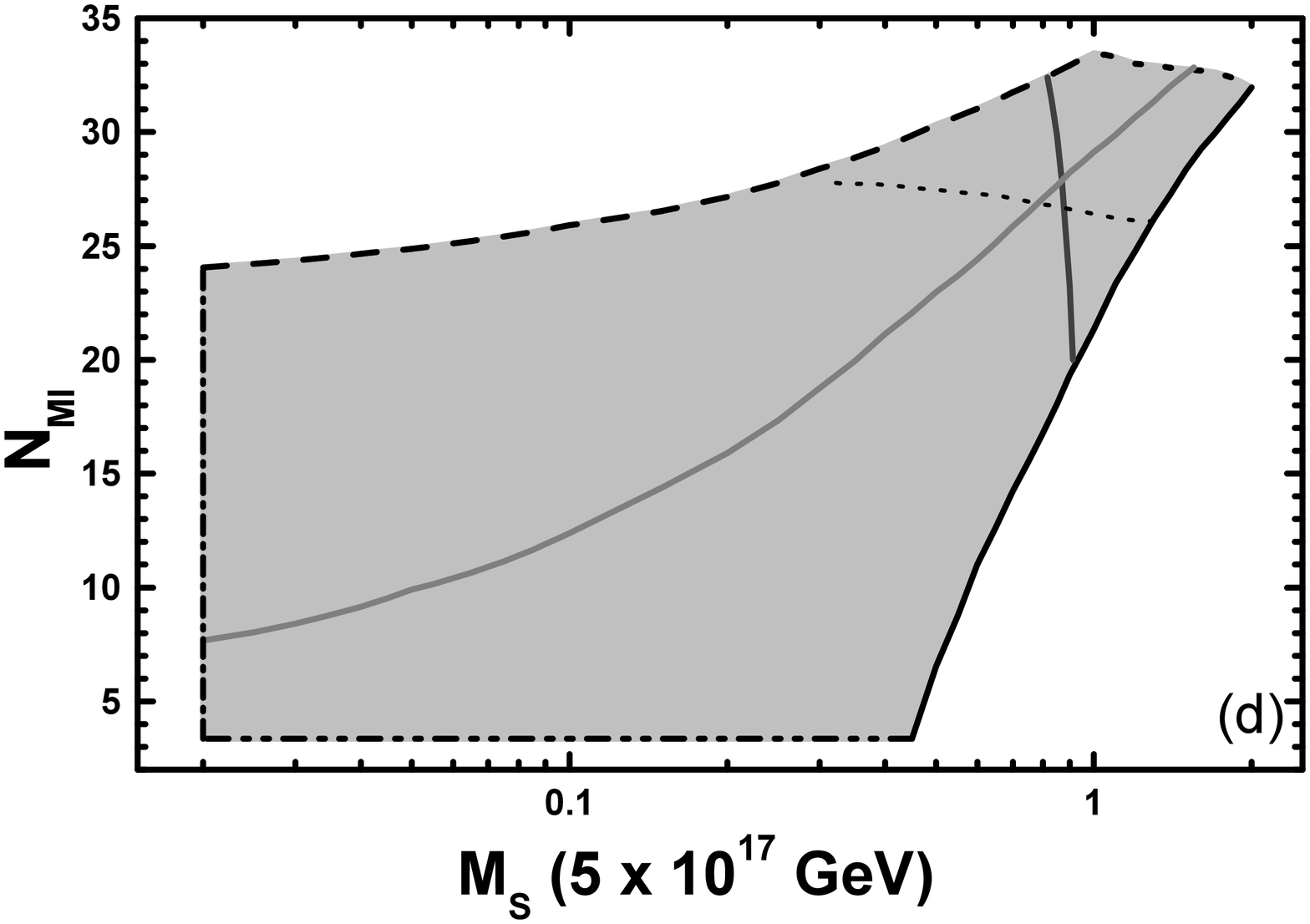}
\caption{\label{smth} Allowed regions in the {(a)}
$M_{\rm S}-v_{_{G}}$, {(b)} $M_{\rm S}-m_s/H_s$, {(c)}
$M_{\rm S}-N_{\rm HI*}$, and {(d)} $M_{\rm S}-N_{\rm MI}$
plane for smooth FHI. The notation is the same as in
Fig.~\ref{shth}. We included small $M_{\rm S}$'s of less
physical interest just to show the effect of the
constraints.}
\end{figure*}

Our results for the three versions of FHI are presented in
Figs.~\ref{stad}$-$\ref{smth}. The conventions adopted for the
various lines are displayed in Table~\ref{tabc}. In
Fig.~\ref{shth}(a) [Fig.~\ref{smth}(a)], we focus on a limited
range of $\kappa$'s [$M_{\rm S}$'s]  for the sake of clarity of
the presentation. Let us discuss each case separately:

\paragraph*{Standard FHI.} In Fig.~\ref{stad}, we present
the regions allowed by Eqs.~(\ref{nswmap}),
(\ref{Prob})--(\ref{nin}), and (\ref{random}) in the {(a)}
$\kappa-v_{_G}$, {(b)}
$\kappa-m_s/H_s$, {(c)} $\kappa-N_{\rm HI*}$, and {(d)}
$\kappa-N_{\rm MI}$ plane for standard FHI. We observe that {(i)}
the resulting $v_{_G}$'s and $\kappa$'s are restricted to rather
large values compared to those allowed within the conventional
(i.e. when $N_{\rm MI}=0$) set-up (compare with
Refs.~\cite{sstad,jp}), {(ii)} as $\kappa$ increases above 0.01
the SUGRA corrections in Eq.~(\ref{Vsugra}) become more and more
significant, (iii) as $\kappa$ decreases below about 0.015 [0.042]
the constraint from the lower [upper] bound on $n_{\rm s}$ in
Eq.~(\ref{nswmap}) ceases to restrict the parameters, since it is
overshadowed by the lower [upper] bound on $N_{\rm HI*}$
[$m_s/H_s$] in Eq.~(\ref{ten}) [Eq.~({\ref{nin})] (indeed, on the
dot-dashed lines $9.84\lesssim N_{\rm HI*}=
N_{\rm HI}^{\rm min}\lesssim10.62$, which implies that
$0.949\gtrsim n_{\rm s}\gtrsim0.926$, while on the double
dot-dashed ones $m_s/H_s\simeq 2.45\Rightarrow N_{\rm MI}
\simeq3.35$ yielding $n_{\rm s}\simeq 0.98-0.99$),
(iv) $|dn_{\rm s}/d\ln k|$ remains well below the
bound in Eq.~(\ref{dnsk}) in the largest part of the
regions allowed by the other constraints, whereas
$-0.005\gtrsim dn_{\rm s}/d\ln k\gtrsim -0.01$ in a very limited
part of these regions, and (v) for
$n_{\rm s}=0.958$, we obtain $0.004\lesssim\kappa\lesssim0.14$,
$0.79\lesssim v_{_G}/( 10^{16}~{\rm GeV})\lesssim1.08$, and
$-0.002\gtrsim dn_{\rm s}/d\ln k\gtrsim-0.01$ as well as $10\lesssim
N_{\rm HI*}\lesssim 21.7$, $35\gtrsim N_{\rm MI} \gtrsim 24$, and
$0.64\lesssim m_s/H_s\lesssim 0.77$.

\begin{table}[t]
\caption{Input and output parameters for our scenario with shifted
($M_{\rm S}=5\times 10^{17}~{\rm GeV}$) or smooth FHI for $n_{\rm
s}=0.958$ and $v_{_G}=M_{\rm GUT}$.}
\begin{tabular}{lr@{\hspace{0.5cm}}clr}
\toprule
\multicolumn{2}{c}{Shifted FHI}&&\multicolumn{2}{c}{Smooth FHI}\\
\colrule\colrule
$\sigma_*~(10^{16}~{\rm GeV})$ & $2.2$&&$\sigma_*~(10^{16}~{\rm
GeV})$ & $23.53$\\
$\kappa$ & $0.01$&&$M_{\rm S}~(5\times 10^{17}~{\rm GeV})$ &
$0.87$\\ \colrule\colrule
$M~(10^{16}~{\rm GeV})$& $2.35$&&$\mu_{\rm S}~(10^{16}~{\rm
GeV})$& $0.188$
\\
$1/\xi$ &  4.54&&$\sigma_{\rm f}~(10^{16}~{\rm GeV})$&$13.42$ \\
$N_{\rm HI*}$ &  $21$&&$N_{\rm HI*}$ & $18$\\
$dn_{\rm s}/d\ln k$ &  $-0.0018$&& $dn_{\rm s}/d\ln k$ &
$-0.0055$\\ \colrule
$N_{\rm MI}$ & $24.3$&&$N_{\rm MI}$ & $27.8$\\
$m_s/H_s$ & $0.77$&&$m_s/H_s$ & $0.72$\\
   \botrule
\end{tabular}
\label{tabsh}
\end{table}

\paragraph*{Shifted FHI.} In Fig.~\ref{shth}, we delineate the
regions allowed by Eqs.~(\ref{nswmap}),
(\ref{Prob})--(\ref{nin}), and (\ref{random}) in the {(a)}
$\kappa-v_{_G}$, {(b)}
$\kappa-m_s/H_s$, {(c)} $\kappa-N_{\rm HI*}$, and {(d)}
$\kappa-N_{\rm MI}$ plane for shifted FHI with $M_{\rm S}=
5\times10^{17}~{\rm GeV}$. We observe that (i) in contrast to the
case of standard FHI, the lower [upper] bound on $N_{\rm HI*}$
[$m_s/H_s$] in Eq.~(\ref{ten})
[Eq.~(\ref{nin})] gives a lower [upper] bound on $v_{_G}$ in the
$\kappa-v_{_G}$ plane, {(ii)} the results on $m_s/H_s$, $N_{\rm
HI*}$, and $N_{\rm MI}$ are quite similar to those for standard
FHI (note that the bounds on $\xi$ do not cut out any
slices of the allowed parameter space), and (iii) $v_{_G}$ comes
out considerably larger than in the case of standard FHI and can
be equal to the SUSY GUT scale (some key inputs and outputs for the
interesting case $v_{_G}=M_{\rm GUT}$ with $n_{\rm s} =0.958$ are
presented in Table~\ref{tabsh}).

\paragraph*{Smooth FHI.} In Fig.~\ref{smth}, we present the regions
allowed by Eqs.~(\ref{nswmap}),
(\ref{Prob})--(\ref{nin}), and (\ref{random}) in the {(a)}
$M_{\rm S}-v_{_G}$, {(b)}
$M_{\rm S}-m_s/H_s$, {(c)} $M_{\rm S}-N_{\rm HI*}$, and {(d)}
$M_{\rm S}-N_{\rm MI}$ plane for smooth FHI. We observe that (i)
the SUGRA corrections in Eq.~(\ref{Vsugra}) play an important role
for every $M_{\rm S}$ in the allowed regions of Fig.~\ref{smth},
{(ii)} in contrast to standard and shifted FHI, $|dn_{\rm s}/d\ln
k|$ is considerably enhanced with $-0.005\gtrsim dn_{\rm s}/d\ln
k\gtrsim -0.01$ holding in a sizable portion of the parameter
space for $v_{_G}\sim M_{\rm GUT}$, (iii) the constraint of
Eq.~(\ref{ten}) does not restrict the parameters unlike the cases
of standard and shifted FHI (on the dashed lines we have $0.02
\lesssim M_{\rm S}/(5\times10^{17}~{\rm GeV})\lesssim1.05$, $12.6
\lesssim N_{\rm HI*}\lesssim21.3$, whereas $N^{\rm min}_{\rm
HI*}\sim10-11$), and (iv) as in the case of shifted FHI, we can
find an acceptable solution fixing $n_{\rm s} =0.958$ and
$v_{_G}=M_{\rm GUT}$ (some key inputs and outputs of this solution
are arranged in Table~\ref{tabsh}).

\section{Conclusions \label{con}}  We investigated a cosmological
scenario tied to two bouts of inflation. The first one is a GUT
scale FHI which reproduces the current data on $P_{\cal R}$ and
$n_{\rm s}$ within the power-law $\Lambda$CDM cosmological model
and generates a limited number of e-foldings $N_{\rm HI*}$. The
second one is an intermediate scale MI which produces the residual
number of e-foldings. We assume that the field which is
responsible for MI is a string axion which remains naturally almost
massless during FHI. We have taken into account extra restrictions
on the parameters originating from (i) the resolution of the
horizon and flatness problems of the standard big bang cosmology,
(ii) the requirements that FHI lasts long enough to generate the
observed primordial fluctuations on all the cosmological scales
and that these scales are not reprocessed by the subsequent MI,
(iii) the limit on the running of $n_{\rm s}$, (iv) the
naturalness of MI, (v) the homogeneity of the present universe,
and (vi) the complete randomization of the string axion during FHI.
Fixing $n_{\rm s}$ to its central value, we concluded that (i)
relatively large $\kappa$'s and $v_{_G}$'s are required within the
standard FHI with $10\lesssim N_{\rm HI*}\lesssim 21.7$ and (ii)
identification of the GUT breaking VEV with the SUSY GUT scale is
possible within shifted [smooth] FHI with $N_{\rm HI*}\simeq 21$
[$N_{\rm HI*}\simeq18$]. In all these cases, MI of the slow-roll
type with $m_s/H_s\sim 0.6-0.8$ and a very mild tuning (of order
0.01) of the initial value of the string axion produces the
necessary additional number of e-foldings. Therefore, MI
complements successfully FHI.

\section*{Acknowledgments}
We would like to thank K. Dimopoulos and R.
Trotta for useful discussions. This work has been supported by the
European Union under the contracts MRTN-CT-2004-503369 and
HPRN-CT-2006-035863 as well as by the PPARC research grant
PP/C504286/1.

\def\ijmp#1#2#3{{Int. Jour. Mod. Phys.}
{\bf #1},~#3~(#2)}
\def\plb#1#2#3{{Phys. Lett. B }{\bf #1},~#3~(#2)}
\def\zpc#1#2#3{{Z. Phys. C }{\bf #1},~#3~(#2)}
\def\prl#1#2#3{{Phys. Rev. Lett.}
{\bf #1},~#3~(#2)}
\def\rmp#1#2#3{{Rev. Mod. Phys.}
{\bf #1},~#3~(#2)}
\def\prep#1#2#3{{Phys. Rep. }{\bf #1},~#3~(#2)}
\def\prd#1#2#3{{Phys. Rev. D }{\bf #1},~#3~(#2)}
\def\npb#1#2#3{{Nucl. Phys. }{\bf B#1},~#3~(#2)}
\def\npps#1#2#3{{Nucl. Phys. B (Proc. Sup.)}
{\bf #1},~#3~(#2)}
\def\mpl#1#2#3{{Mod. Phys. Lett.}
{\bf #1},~#3~(#2)}
\def\arnps#1#2#3{{Annu. Rev. Nucl. Part. Sci.}
{\bf #1},~#3~(#2)}
\def\sjnp#1#2#3{{Sov. J. Nucl. Phys.}
{\bf #1},~#3~(#2)}
\def\jetp#1#2#3{{JETP Lett. }{\bf #1},~#3~(#2)}
\def\app#1#2#3{{Acta Phys. Polon.}
{\bf #1},~#3~(#2)}
\def\rnc#1#2#3{{Riv. Nuovo Cim.}
{\bf #1},~#3~(#2)}
\def\ap#1#2#3{{Ann. Phys. }{\bf #1},~#3~(#2)}
\def\ptp#1#2#3{{Prog. Theor. Phys.}
{\bf #1},~#3~(#2)}
\def\apjl#1#2#3{{Astrophys. J. Lett.}
{\bf #1},~#3~(#2)}
\def\n#1#2#3{{Nature }{\bf #1},~#3~(#2)}
\def\apj#1#2#3{{Astrophys. J.}
{\bf #1},~#3~(#2)}
\def\anj#1#2#3{{Astron. J. }{\bf #1},~#3~(#2)}
\def\mnras#1#2#3{{MNRAS }{\bf #1},~#3~(#2)}
\def\grg#1#2#3{{Gen. Rel. Grav.}
{\bf #1},~#3~(#2)}
\def\s#1#2#3{{Science }{\bf #1},~#3~(#2)}
\def\baas#1#2#3{{Bull. Am. Astron. Soc.}
{\bf #1},~#3~(#2)}
\def\ibid#1#2#3{{\it ibid. }{\bf #1},~#3~(#2)}
\def\cpc#1#2#3{{Comput. Phys. Commun.}
{\bf #1},~#3~(#2)}
\def\astp#1#2#3{{Astropart. Phys.}
{\bf #1},~#3~(#2)}
\def\epjc#1#2#3{{Eur. Phys. J. C}
{\bf #1},~#3~(#2)}
\def\nima#1#2#3{{Nucl. Instrum. Meth. A}
{\bf #1},~#3~(#2)}
\def\jhep#1#2#3{{J. High Energy Phys.}
{\bf #1},~#3~(#2)}

\end{document}